\documentclass[12pt,fleqn]{article}
\usepackage{amsfonts}
\newtheorem{thm}{\sc Theorem}[section]

\newtheorem{lemma}[thm]{\sc Lemma}
\newtheorem{cor}[thm]{\sc Corollary}
\newtheorem{conj}[thm]{\sc Conjecture}
\newcommand{\lipic}{\mbox{
\unitlength 0.5mm
\begin{picture}(8.00,8.00)(0,2.00)
\put(4.00,4.00){\circle{8.00}}
\put(4.00,6.00){\line(0,-1){4.00}}
\end{picture}
}}
\newcommand{\wepic}{\mbox{
\unitlength 0.5mm
\begin{picture}(8.00,8.00)(0,2.00)
\put(4.00,4.00){\circle{8.00}}
\put(2.00,6.00){\line(1,0){4.00}}
\put(2.00,6.00){\line(0,-1){4.00}}
\end{picture}
}}
\newcommand{\papic}{\mbox{
\unitlength 0.5mm
\begin{picture}(8.00,8.00)(0,2.00)
\put(4.00,4.00){\circle{8.00}}
\put(2.00,6.00){\line(0,-1){4.00}}
\put(6.00,6.00){\line(0,-1){4.00}}
\end{picture}
}}
\newcommand{\trpic}{\mbox{
\unitlength 0.5mm
\begin{picture}(8.00,8.00)(0,2.00)
\put(4.00,4.00){\circle{8.00}}
\put(1.00,5.00){\line(1,0){6.00}}
\put(4.00,5.00){\line(0,-1){4.00}}
\end{picture}
}}
\def\blob{\quad\rule{5pt}{5pt}}  
\def\la{\langle}
\def\ra{\rangle}

\def\ie{\emph{i.e.}}
\def\eg{\emph{e.g.}}
\def\eq{\begin{equation}}
\def\en{\end{equation}}
\def\eqa{\begin{eqnarray}}
\def\ena{\end{eqnarray}}
\def\nn{\nonumber \\}
\def\ot{\otimes}

\def\12{\frac{1}{2}}

\def\C{\mathbb{C}} 
\def\Re{\mathbb{R}} 
\def\N{\mathbb{N}} 
\def\R{{\cal R}} 
\def\H{{\cal H}}  
\def\U{\!\uparrow} 
\def\D{\!\downarrow} 
\def\tr{\mbox{tr}\,} 
\def\ep{\epsilon}
\def\al{\alpha}
\def\be{\beta}
\def\ga{\gamma}
\def\om{\omega}
\def\Om{\Omega}
\def\dOm{\frac{d\Om}{4\pi}}
\begin{document}
\begin{titlepage}
\rightline{Feb.\ 9, 1999}
\rightline{PUPT-1832}
\vspace{2em}
\begin{center}{\bf\Large On Lieb's conjecture for the 
Wehrl entropy of Bloch coherent states}\\[2em]
Peter Schupp\\[2em]
{\sl Department of Physics,
Princeton University\\Princeton, NJ 08544-0708, USA} \\[6em]
\end{center}
\begin{abstract}
Lieb's conjecture for the Wehrl entropy of Bloch coherent states
is proved for spin 1 and spin 3/2. Using a geometric representation
we solve the entropy integrals for states of arbitrary spin
and evaluate them explicitly in the cases of
spin 1, 3/2, and 2. We also give a group theoretic
proof for all spin of a related inequality.
\end{abstract}

\vfill
\noindent \hrule
\vskip.2cm
\noindent{\tiny 
\copyright 1999 by the author. 
Reproduction of this article, in its entirety, by any
means is permitted for non-commercial purposes.}
\hbox{{\small{\it e-mail: }}{\small \quad schupp@princeton.edu}}
\end{titlepage}
%
\setcounter{page}{1}

\section{Introduction}

Wehrl proposed \cite{W} 
a hybrid between quantum mechanical and classical entropy
that enjoys monotonicity, strong subadditivity and positivity -- physically
desirable properties, some of which both kinds of entropy lack \cite{B,E}. 
This new entropy is the ordinary Shannon entropy of
the probability density provided by the lower
symbol of the density matrix. 

For a quantum mechanical system with density matrix $\rho$, Hilbert space
$\H$,
and a family of normalized
coherent states $|z\ra$,
parametrized symbolically by
$z$ and satisfying
$\int dz \, |z\ra\la z| = {\bf 1}$ (resolution of identity),
the Wehrl entropy is
\eq
S_W(\rho)  = -\int dz \, \la z|\rho|z\ra \ln \la z|\rho|z\ra . \label{W0}
\en
%
%
Like quantum mechanical entropy,
$S_Q = -\tr \rho\ln\rho $,
Wehrl entropy is always non-negative, in fact
$S_W > S_Q \geq 0$.
In view of this inequality it is interesting to ask for the minimum
of $S_W$ and the corresponding minimizing density matrix.
It follows from concavity of $-x \ln x$ that a minimizing
density matrix must be a pure state, \ie, $\rho = |\psi\ra\la\psi|$
for a normalized vector $|\psi\ra \in \H$ \cite{A}. (Note that
$S_W(|\psi\ra\la\psi|)$ depends on $|\psi\ra$ and is non-zero,
unlike the quantum entropy which is of course zero for pure states.)

For Glauber coherent states  Wehrl conjectured \cite{W} and Lieb proved
\cite{A}
that the minimizing state $|\psi\ra$ is again a coherent
state. It turns out that all Glauber coherent states
have Wehrl entropy one, so Wehrl's conjecture can be written as follows:
\begin{thm}[\rm Lieb] \label{thmWL}
The minimum of $S_W(\rho)$ for states in $\H = L^2(\Re)$
is one,
\eq
S_W(|\psi\ra\la\psi|)
= -\int dz \, |\la\psi|z\ra|^2 \ln |\la\psi|z\ra|^2 \geq 1 , \label{W1}
\en
with equality if and only if $|\psi\ra$ is a coherent state.
\end{thm}
To prove this, Lieb used a clever combination of the sharp
Hausdorff-Young inequality \cite{Y1,Y3,LL}
and the sharp Young inequality \cite{Y2,Y1,Y3,LL}
to show that
\eq
s \int dz \, |\la z|\psi\ra|^{2 s} \leq 1, \quad s \geq 1, \label{W2}
\en
again with equality if and only if $|\psi\ra$ is a coherent state.
Wehrl's conjecture follows from this in the limit $s \rightarrow 1$
essentially because
(\ref{W1}) is the derivative of (\ref{W2}) with respect to $s$ at $s=1$.
All this easily generalizes to $L^2(\Re^n)$ \cite{A,Y4}.

The lower bound on the Wehrl entropy is related to
Heisenberg's uncertainty principle \cite{AH,G} and it has been speculated that
$S_W$ can be used to measure uncertainty due to both quantum and thermal
fluctuations \cite{G}.

It is very surprising that `heavy artillery' like the sharp constants
in the mentioned inequalities are needed in Lieb's proof. To elucidate
this situation, Lieb suggested \cite{A} studying the analog of Wehrl's
conjecture for Bloch coherent states $|\Omega\ra$, where one should
expect significant 
simplification since these are finite dimensional Hilbert spaces.
However, no progress has been made, not even for a single spin, even though
many attempts have been made \cite{B}. Attempts to proceed again 
along the lines of Lieb's original proof have failed to provide a sharp
inequality and the direct computation of the entropy and related integrals,
even numerically, was unsuccesful \cite{S}.

The key to the recent progress is a geometric representation of a
state of spin~$j$ as $2j$ points on a sphere.
In this representation the expression
$|\la\Omega|\psi\ra|^2$ factorizes into a product
of $2j$ functions $f_i$ on the sphere,  
which measure the square chordal distance
from the antipode of the point parametrized by
$\Omega$ to each of the $2j$ points on the sphere.
Lieb's conjecture,
in a generalized form analogous to (\ref{W2}),
then looks like the quotient of two H\"older inequalities
\eq
\frac{|\!|f_1 \cdots f_{2j}|\!|_s}{|\!|f_1 \cdots f_{2j}|\!|_1}
\leq 
\frac{\prod_{i=1}^{2j}|\!|f_i|\!|_{2js}}{\prod_{i=1}^{2j} |\!|f_i|\!|_{2j}},
\label{holder}
\en
with the one with the higher power winning against the other one.
We shall give a group theoretic proof of this inequality
for the special case $s \in \N$ in theorem~\ref{natural}.

In the geometric representation the 
Wehrl entropy of spin states finds a direct physical
interpretation: It is the classical entropy of a single particle on a sphere
interacting via Coulomb potential with $2j$ fixed sources; $s$ plays the
role of inverse temperature.

The entropy integral (\ref{W0}) can now be done because 
$|\la\Omega|\psi\ra|^2$ factorizes
and one finds a formula for the Wehrl entropy of any state.
When we evaluate the entropy explicitly for states of spin 1,
3/2, and 2 we find surprisingly simple expressions  solely in
terms of the square chordal distances between the points on the
sphere that define the given state. 

A different, more group theoretic approach seems to point 
to a connection between
Lieb's conjecture and the norm of certain spin $j s$ states with
$1 \leq s \in \Re$ \cite{J}.
So far, however, this has only been useful for proving the analog
of inequality (\ref{W2}) for $s \in \N$.

We find that a proof of Lieb's conjecture for low spins can be reduced
to some beautiful spherical geometry, 
but the unreasonable difficulty of a complete proof
is still a great puzzle; its resolution may very well lead to
interesting mathematics and perhaps physics.

\section{Bloch coherent spin states}

Glauber coherent states 
$|z\ra = \pi^{-\frac{1}{4}} e^{-(x-q)^2/2} e^{ipx}$,
parametrized by $z = (q + i p)/\sqrt{2}$ and with measure
$dz = dp dq/2\pi$,
are usually introduced as
eigenvectors of the annihilation operator
$a = (\hat x + i \hat p)/\sqrt{2}$, 
$a|z\ra = z |z\ra$,
but the same states can also be
obtained by the action of the Heisenberg-Weyl group
$H_4 = \{a^\dagger a, a^\dagger, a, I\}$
on the extremal state
$|0\ra = \pi^{-\frac{1}{4}} e^{-x^2/2}$.
Glauber coherent states are thus elements of the coset
space of the Heisenberg-Weyl group
modulo the stability subgroup $U(1)\ot U(1)$ that leaves the extremal state
invariant. (See \eg\ \cite{C1} and references therein.)
This construction easily generalizes
to other groups, in particular to SU(2), where it gives
the Bloch coherent spin states \cite{BC} that we are interested in:
Here the Hilbert space can be any one of the finite dimensional spin-$j$
representations $[j] \equiv \C^{2j+1}$ of SU(2), 
$j = {1\over 2}, 1, \frac{3}{2}, \ldots$,
and
the extremal state for each $[j]$ is the
highest weight vector $|j,j\ra$. The stability subgroup is U(1)
and the coherent states are thus 
elements of the sphere $S_2 = $SU(2)/U(1);
they can be labeled by
$\Omega = (\theta,\phi)$ and are
obtained from $|j,j\ra$ by rotation:
\eq
|\Omega\ra_j = \R_j(\Omega) |j,j\ra. \label{Om}
\en
For spin $j = \12$ we find
\eq
|\omega\ra = p^{\12} e^{-i{\phi\over 2}} |\U\ra 
           + (1-p)^{\12} e^{i{\phi\over 2}} |\D\ra ,  \label{coh}
\en
with $p \equiv \cos^2\frac{\theta}{2}$.
(Here and in the following $|\omega\ra$ is short for the spin-$\12$ coherent
state $|\Omega\ra_\12$; $\omega = \Omega = (\theta,\phi)$. 
$|\U\ra \equiv |\12,\12\ra$ and
$|\D\ra \equiv |\12,-\12\ra$.)
An important observation for what follows is that
the product of two coherent states for the same $\Omega$
is again a coherent state:
\eqa 
|\Omega\ra_j \ot |\Omega\ra_{j'} 
	& = & (\R_j \ot \R_{j'})\, (|j,j\ra \ot |j',j'\ra)  \nn
	& = & \R_{j+j'}\, |j+j',j+j'\ra 	           
	\; = \; |\Omega\ra_{j+j'} .	     
\ena
Coherent states are in fact the only states for which
the product of a spin-$j$ state with a spin-$j'$ state is
a spin-$(j+j')$ state and not a more general element of
$[j+j'] \oplus \ldots \oplus [\,|j - j'|\,]$.
From this key property
an explicit representation for Bloch coherent states of higher spin 
can be easily derived:
\eqa
|\Omega\ra_j & = & \left(|\omega\ra\right)^{\ot 2j} 
             \; = \;  \left(p^{\12} e^{-i{\phi\over 2}} |\U\ra 
               + (1-p)^{\12} e^{i{\phi\over 2}} |\D\ra\right)^{\ot 2j} \nn
	     & = & \sum_{m=-j}^j {2 j \choose j + m}^{\12}
	     p^{j+m\over 2}  (1-p)^{j-m\over 2} 
	     e^{-i m {\phi\over 2}} |j,m\ra.  \label{Coh}
\ena
(The same expression can also be obtained directly from (\ref{Om}), see
\eg\ \cite[chapter 4]{C2}.)
The coherent states as given are normalized $\la\Om|\Om\ra_j =1$
and satisfy
\eq
(2j+1) \int\dOm \, |\Om\ra_j\la\Om|_j = P_j , \qquad \mbox{(resolution of
identity)} \label{project}
\en
where $P_j = \sum |j,m\ra\la j,m|$ is the projector onto $[j]$.
It is not hard to compute the Wehrl entropy for a coherent state
$|\Om'\ra$: Since the integral over the sphere is invariant under rotations
it is enough to consider the coherent state $|j,j\ra$; then use
$|\la j,j|\Om\ra|^2 = |\la\U\!\!|\om\ra|^{2\cdot 2j} = p^{2j}$
and $d\Om/4\pi = -dp\,d\phi/2\pi$, where 
$p =\cos^2 \frac{\theta}{2}$ as above, to obtain
\eqa
S_W(|\Om'\ra\la\Om'|) 
& = & -(2j+1) \int\dOm \, |\la\Om|\Om'\ra|^2 \ln |\la\Om|\Om'\ra|^2 \nn
& = & -(2j+1) \int_0^1 dp \, p^{2j} \, 2j \ln p \, = \, \frac{2j}{2j+1}.
\ena
Similarly, for later use,
\eq
(2js+1) \int \dOm |\la\Om'|\Om\ra|^{2s} = (2js+1) \int_0^1 dp \, p^{2js} = 1.
\en
As before the density matrix that minimizes $S_W$
must be a pure state $|\psi\ra\la\psi|$. 
The analog of theorem~\ref{thmWL} for spin states is:
\begin{conj}[\rm Lieb] \label{conject1}
The minimum of $S_W$ for states in $\H = \C^{2j+1}$
is $2j/(2j+1)$,
\eq
S_W(|\psi\ra\la\psi|) 
= -(2j+1) \int\dOm \, |\la\Om|\psi\ra|^2 \ln |\la\Om|\psi\ra|^2
\geq \frac{2j}{2j+1},
\en
with equality if and only if $|\psi\ra$ is a coherent state.
\end{conj}

\noindent
\emph{Remark:} For spin 1/2 this is an identity because
all spin~1/2 states are coherent states. The first non-trivial case
is spin $j=1$.

\section{Proof of Lieb's conjecture for low spin}

In this section we shall geometrize the description of spin states, use this
to solve the entropy integrals
for all spin and prove Lieb's conjecture for low spin by actual computation
of the entropy.

\begin{lemma}
States of spin $j$ are in one to one correspondence to $2j$ points
on the sphere $S_2$: With $2j$ points,
parametrized by $\om_k = (\theta_k, \phi_k)$, $k = 1, \ldots , 2j$,
we can associate a state
\eq
|\psi\ra = c^\12 P_j (|\om_1\ra \ot \ldots \ot |\om_{2j}\ra) \; \in \; [j] ,
\label{psiprod}
\en
and every state $|\psi\ra \in [j]$ is of that form. (The spin-$\12$ states
$|\om_k\ra$ are given by (\ref{coh}),
$c^\12 \neq 0$ fixes the
normalization of $|\psi\ra$, and $P_j$ is the projector onto spin~$j$.)
\label{sphere}
\end{lemma}

\noindent \emph{Remark:} 
Some or all of the points may coincide.
Coherent states are exactly
those states for which all points on the sphere coincide. 
$c^\12 \in \C$ may contain an (unimportant) phase
that we can safely ignore in the following.
This representation is unique up to permutation
of the $|\om_k\ra$. The $\om_k$ may be found by looking at
$\la\Om|\psi\ra$ as a function of $\Om = (\theta,\psi)$:
they are the antipodal points to the zeroes of this function. 

\noindent {\sc Proof}:
Rewrite (\ref{Coh}) in complex coordinates for $\theta\neq 0$
\eq
z = \left(\frac{p}{1-p}\right)^\12 e^{i\phi} = \cot\frac{\theta}{2} e^{i\phi}
\en
(stereographic projection)
and contract it with $|\psi\ra$ to find
\eq
\la\Omega|\psi\ra = \frac{e^{-ij\phi}}{(1+z\bar z)^j} 
\sum_{m=-j}^{j_{\mbox{\tiny max}}} 
	   {2 j \choose j + m}^{\12} z^{j+m} \psi_m , \label{poly}
\en
where $j_{\mbox{\scriptsize max}}$ is the largest value of $m$ for which
$\psi_m$ in the expansion
$|\psi\ra = \sum\psi_m |m\ra$
is nonzero. This is a polynomial of degree
$j+j_{\mbox{\scriptsize max}}$ in $z \in \C$ and can thus be factorized:
\eq
\la\Omega|\psi\ra = 
\frac{ e^{-ij\phi} \psi_{j_{\mbox{\tiny max}}} }{ (1+z\bar z)^j } 
\prod_{k=1}^{j+j_{\mbox{\tiny max}}} (z - z_k) . \label{fact}
\en
Consider now the spin~$\12$ states
$|\om_k\ra = (1+z_k \bar z_k)^{-\12}(|\U\ra - z_k |\D\ra)$
for $1 \leq k \leq j+j_{\mbox{\tiny max}}$ and
$|\om_m\ra = |\D\ra$ for $j+j_{\mbox{\tiny max}} < m \leq 2j$. According
to (\ref{poly}):
\eq
\la\om|\om_k\ra = \frac{e^{-\frac{i\phi}{2}}}{(1+z\bar z)^\12 
(1+ z_k\bar z_k)^\12}(z - z_k) , 
\qquad \la\om|\om_m\ra = 
\frac{e^{-\frac{i\phi}{2}}}{(1+z\bar z)^\12},
\en
so by comparison with (\ref{fact}) and with an appropriate constant
$c$
\eq
\la\Om|\psi\ra 
= c^\12 \la\om|\om_1\ra \cdots \la\om|\om_{2j}\ra
= c^\12 \la\Om|\om_1\ot\ldots\ot\om_{2j}\ra.
\label{fac}
\en
By inspection we see that this expression is still valid when
$\theta = 0$ and with the help of (\ref{project}) we can complete the
proof the lemma.$\blob$\\[1ex]
We see that the geometric representation of spin states leads to a
factorization of $\la\Om|\psi\ra|^2$. In this representation we can
now do the entropy integrals, essentially because the logarithm becomes a
simple sum.

\begin{thm}   \label{theorem}
Consider any state $|\psi\ra$ of spin $j$. According to
lemma~\ref{sphere}, it can be written as
$|\psi\ra = c^\12 P_j (|\om_1\ra \ot \ldots \ot |\om_{2j}\ra).$
Let $\R_i$ be the rotation that turns $\om_i$ to the `north pole',
$\R_i|\om_i\ra = |\U\ra$, let $|\psi^{(i)}\ra = \R_i|\psi\ra$,
and let $\psi_m^{(i)}$ be the coefficient of $|j,m\ra$ in the expansion
of $|\psi^{(i)}\ra$,
then the Wehrl entropy is:
\eq
S_W(|\psi\ra\la\psi|) =
\sum_{i=1}^{2j} \sum_{m=-j}^{j} \left(\sum_{n=0}^{j-m}
\frac{1}{2j+1-n}\right) |\psi_m^{(i)}|^2  -  \ln c . \label{formula}
\en
\end{thm}

\noindent \emph{Remark:}
This formula reduces the computation of the Wehrl entropy of any
spin state to its factorization in the sense of lemma~\ref{sphere},
which in general requires the solution of a
$2j$'th order algebraic equation. This may explain why previous 
attempts to do the entropy integrals have failed.
The $n=0$ terms in the expression for the entropy
sum up to $2j/(2j+1)$, the entropy of a coherent state, 
and Lieb's conjecture can be thus be written
\eq
\ln c \leq \sum_{i=1}^{2j} \sum_{m=-j+1}^{j-1} \left(\sum_{n=1}^{j-m}
\frac{1}{2j+1-n}\right) |\psi_m^{(i)}|^2.
\en
Note that $\psi^{(i)}_{-j} = 0$ by construction of
$|\psi^{(i)}\ra$: $\psi^{(i)}_{-j}$ contains a factor
$\la\downarrow|\U\ra$.\\
A similar calculation gives
\eq
\ln c = 2j + \int\dOm \, \ln|\la\Om|\psi\ra|^2 .
\en

\noindent
{\sc Proof}:
Using lemma~\ref{sphere}, (\ref{project}),
the rotational invariance of the
measure and the inverse Fourier transform in $\phi$ we find
\eqa
\lefteqn{S_W(|\psi\ra\la\psi|) \; = \;
{ -(2j+1)}
\int\dOm |\la\Om|\psi\ra|^2 \sum_{i=1}^{2j} \ln |\la\om|\om_i\ra|^2
- \ln c} \nn
&& = { -(2j+1)} \sum_{i=1}^{2j} \int\dOm
|\la\Om|\psi^{(i)}\ra|^2  \ln |\la\om|\U\ra|^2  - \ln c \nn
&& = {\scriptstyle -(2j+1)} \sum_{i=1}^{2j}\sum_{m=-j}^j |\psi^{(i)}_m|^2
{\scriptstyle {2j \choose j + m}}
\int_0^1 dp \, 
 p^{j+m} (1 \! - \! p)^{j-m} \ln p - \ln c.
\ena
It is now easy to do the remaining $p$-integral by partial integration
to proof the theorem.$\blob$\\[1ex]
Lieb's conjecture for low spin can be proved with the help of
formula (\ref{formula}). For spin 1/2 there is
nothing to prove, since all states  of spin 1/2 are coherent states.
The first nontrivial case is spin 1:

\begin{cor}[\rm spin 1]
Consider an arbitrary state of spin 1. Let
$\mu$ be the square of the
chordal distance between the two points on the sphere of radius~$\12$
that represent this state. It's Wehrl entropy is given by
\eq
S_W(\mu) = \frac{2}{3} + c\cdot\left(\frac{\mu}{2} + \frac{1}{c} \ln
\frac{1}{c}\right) , \label{entropy1}
\en
with
\eq
\frac{1}{c} = 1 - \frac{\mu}{2}.
\en
Lieb's conjecture holds for all states of spin 1:
$S_W(\mu) \geq 2/3 = 2j/(2j+1)$ with equality for $\mu = 0$, \ie\ for
coherent states.
\end{cor}

\noindent {\sc Proof}:
Because of rotational invariance we can assume without loss of generality
that the first point is at the `north pole' of the sphere and that
the second point is parametrized as $\om_2 = (\tilde\theta, \tilde\phi = 0)$,
so that
$\mu = \sin^2\frac{\tilde\theta}{2}$ . Up to normalization (and an irrelevant
phase)
\eq
|\tilde\psi\ra = P_{j=1}|\U\ot\tilde\om\ra
\en
is the state of interest. But from (\ref{coh})
\eq
|\U\ot\tilde\om\ra = (1-\mu)^\12|\U\;\U\ra + \mu^\12|\U\;\D\ra.
\en
Projecting onto spin 1 and inserting the normalization constant $c^\12$
we find
\eq
|\psi\ra = c^\12\left((1-\mu)^\12 |1,1\ra + \mu^\12
\frac{1}{\sqrt{2}}|1,0\ra\right). \label{state}
\en
This gives (ignoring a possible phase) 
\eq
1 = \la\psi|\psi\ra = c\left(1 - \mu + \frac{\mu}{2}\right) = c\left(1 -
\frac{\mu}{2}\right)   \label{cvalue}
\en
and so $1/c = 1 - \mu/2$. Now we need to compute the components
of $|\psi^{(1)}\ra$ and $|\psi^{(2)}\ra$. Note that
$|\psi^{(1)}\ra = |\psi\ra$ because $\om_1$ is already pointing to
the `north pole'. To obtain $|\psi^{(2)}\ra$ we need to rotate point 2
to the `north pole'. We can use the remaining rotational freedom
to effectively exchange the two points, thereby recovering the original
state $|\psi\ra$. The components of 
both $|\psi^{(1)}\ra$ and $|\psi^{(2)}\ra$ can thus be read off (\ref{state}):
\eq
\psi_1^{(1)} = \psi_1^{(2)} = c^\12(1-\mu)^\12 ,
\qquad \psi_0^{(1)} = \psi_0^{(2)} = c^\12 \mu^\12/\sqrt{2}.
\en
Inserting now $c$, $|\psi_1^{(1)}|^2 = |\psi_1^{(2)}|^2 = c(1-\mu)$,
and $|\psi_0^{(1)}|^2 = |\psi_0^{(2)}|^2 = c \mu/2$ into (\ref{formula})
gives the stated entropy.

To prove Lieb's conjecture for states of spin~1 we use
(\ref{cvalue}) to show that
the second term in (\ref{entropy1}) is always non-negative and zero only for
$\mu = 0$, \ie\ for a coherent state. This follows from
\eq
\frac{c \mu}{2} - \ln c \geq \frac{c \mu}{2} + 1 - c = 0
\en
with equality for $c=1$ which is equivalent to $\mu=0$ .$\blob$
\vspace{2ex}
\begin{figure}
\begin{center}
\unitlength 1.00mm
\linethickness{0.4pt}
\begin{picture}(30.00,21.00)(10,5)
\put(10.00,0.00){\line(5,1){25.00}}
\put(35.00,5.00){\line(-4,3){20.00}}
\put(15.00,20.00){\line(-1,-4){5.00}}
\put(9.00,0.00){\makebox(0,0)[rc]{$1$}}
\put(15.00,21.00){\makebox(0,0)[cb]{$3$}}
\put(36.00,5.00){\makebox(0,0)[lc]{$2$}}
\put(11.00,10.00){\makebox(0,0)[rc]{$\mu$}}
\put(26.00,13.00){\makebox(0,0)[lb]{$\epsilon$}}
\put(27.00,2.00){\makebox(0,0)[ct]{$\nu$}}
\end{picture}
\end{center}
\caption{Spin~3/2}
\label{three}
\end{figure}
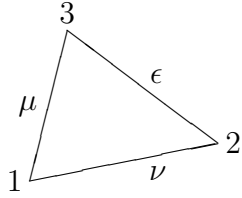
\begin{cor}[\rm spin 3/2]
Consider an arbitrary state of spin 3/2. Let $\ep$, $\mu$, $\nu$
be the squares of the chordal distances between the three points on
the sphere of radius $\12$ that represent this state (see figure~\ref{three}). 
It's Wehrl entropy
is given by
\eq
S_W(\ep,\mu,\nu) = \frac{3}{4} + 
c\cdot\left(\frac{\ep+\mu+\nu}{3} - \frac{\ep\mu + \ep\nu + \mu\nu}{6}
+\frac{1}{c} \ln\frac{1}{c} \right)  \label{spin32}
\en
with 
\eq
\frac{1}{c} = 1 - \frac{\ep+\mu+\nu}{3} .
\en
Lieb's conjecture holds for all states of spin 3/2:
$S_W(\ep,\mu,\nu) \geq 3/4 = 2j/(2j+1)$ with equality for 
$\ep = \mu = \nu = 0$, \ie\ for
coherent states.
\end{cor}

\noindent {\sc Proof}: The proof is similar to the spin~1 case, but the
geometry and algebra is more involved.
Consider a sphere of radius ${1\over 2}$, with points 1, 2, 3 on its surface,
and two planes through its center; the first plane
containing
points 1 and 3, the second plane containing points 2 and 3. The intersection
angle $\phi$
of these two planes satisfies
\eq
 2\cos\phi \sqrt{\ep\mu(1-\ep)(1-\mu)} = \ep + \mu - \nu - 2\ep\mu .
\label{phi}
\en
$\phi$ is the azimuthal angle of point 2, if point 3 is at the `north pole' of
the sphere and point 1 is assigned zero azimuthal angle.

The states $|\psi^{(1)}\ra$,
$|\psi^{(2)}\ra$, and $|\psi^{(3)}\ra$ all have one point at the
north pole of the sphere. It is enough to compute the values of
$|\psi_m^{(i)}|^2$ for
one $i$, the other values can be found by appropriate permutation of
$\ep$, $\mu$, $\nu$. (Note that we make no restriction on
the parameters $0\leq \ep$, $\mu$, $\nu \leq 1$ other than that they are
square chordal distances between three points on a sphere of
radius $\12$.)
We shall start with $i = 3$: Without loss of generality
the three points can be parametrized as $\om^{(3)}_1 = (\tilde\theta,0)$,
$\om^{(3)}_2 = (\theta,\phi)$, and $\om^{(3)}_3 = (0,0)$
with $\mu = \sin^2{\tilde\theta\over 2}$ and $\ep = \sin^2{\theta\over 2}$.
Corresponding spin-$\12$ states are
\eqa
|\om^{(3)}_1\ra & = & (1-\mu)^\12|\U\ra + \mu^\12|\D\ra ,\label{om1}\\
|\om^{(3)}_2\ra & = & (1-\ep)^\12 e^{-i\phi\over 2}|\U\ra 
+ \ep^\12 e^{i\phi\over 2}|\D\ra ,  \label{om2}\\
|\om^{(3)}_3\ra & = & |\U\ra ,      \label{om3} 
\ena
and up to normalization, the state of interest is
\eqa
|\tilde\psi^{(3)}\ra 
& = & P_{j=3/2} |\om^{(3)}_1\ot\om^{(3)}_2\ot\om^{(3)}_3\ra  \nn
& = & (1-\ep)^\12 (1-\mu)^\12 e^{-i\phi\over 2}
      |{3\over 2},{3\over 2}\ra \nn
&&    + \left( (1-\mu)^\12 \ep^\12 e^{i\phi\over 2} 
      + \mu^\12 (1-\ep)^\12 e^{-i\phi\over 2} \right) 
      { {1 \over \sqrt{3}}} |{3\over 2},{1\over 2}\ra \nn
&&    + \mu^\12 \ep^\12 e^{i\phi\over 2} 
      { {1 \over \sqrt{3}}} |{3\over 2},-{1\over 2}\ra .
\ena
This gives 
\eqa
|\tilde\psi^{(3)}_{3\over 2}|^2 & = & (1-\ep)(1-\mu),\\
|\tilde\psi^{(3)}_{1\over 2}|^2 & = & {1 \over 3}\left(
\ep(1-\mu) + \mu(1 - \ep) + 2 \sqrt{\ep\mu(1-\mu)(1-\ep)} \cos\phi\right) \nn
& = & {2\over 3}\ep(1-\mu) + {2\over 3}\mu(1 - \ep) -{\nu\over 3}, \\
|\tilde\psi^{(3)}_{-{1\over 2}}|^2 & = & {\ep \mu\over 3},
\ena
and
$|\tilde\psi^{(3)}_{-{3\over 2}}|^2  = 0$. The sum of these expressions
is
\eq
{1\over c} = \la\tilde\psi|\tilde\psi\ra =
1 - {\ep + \mu + \nu \over 3} ,
\en
with $0 < 1/c \leq 1$.
The case $i=1$ is found by exchanging $\mu \leftrightarrow \nu$ (and also
$3 \leftrightarrow 1$, $\phi \leftrightarrow -\phi$).
The case $i=2$ is found by permuting
$\ep\rightarrow\mu\rightarrow\nu\rightarrow\ep$ (and also $1 \rightarrow 3
\rightarrow 2 \rightarrow 1$).
Using (\ref{formula}) then gives the stated entropy.

To complete the proof Lieb's conjecture for all states of 
spin~$3/2$ we need to show
that the second term in (\ref{spin32}) is always non-negative and zero
only for $\ep=\mu=\nu=0$.
From the inequality $(1-x)\ln(1-x) \geq -x + x^2/2$ for $0 \leq x < 1$,
we find
\eq
{1\over c}\ln{1\over c} \geq -{\ep+\mu+\nu\over 3} + {1\over 2}\left(
{\ep+\mu+\nu\over 3}\right)^2 ,
\en
with equality for $c=1$. Using the inequality between algebraic and geometric
mean it is not hard to see that
\eq
\left({\ep+\mu+\nu\over 3}\right)^2 \geq {\ep\mu + \nu\ep + \mu\nu \over 3}
\en
with equality for $\ep=\mu=\nu$. Putting everything together and inserting
it into (\ref{spin32}) we have, as desired, $S_W \geq 3/4$ with equality
for $\ep=\mu=\nu=0$, \ie\ for coherent states.$\blob$
\vspace{1ex}
\begin{figure}
\begin{center}
\unitlength 1.00mm
\linethickness{0.4pt}
\begin{picture}(30.00,36.00)(10,7)
\put(10.00,15.00){\line(5,1){25.00}}
\put(35.00,20.00){\line(-4,3){20.00}}
\put(15.00,35.00){\line(-1,-4){5.00}}
\put(10.00,15.00){\line(3,-2){15.00}}
\put(25.00,5.00){\line(2,3){10.00}}
\put(9.00,15.00){\makebox(0,0)[rc]{$1$}}
\put(15.00,36.00){\makebox(0,0)[cb]{$3$}}
\put(36.00,20.00){\makebox(0,0)[lc]{$2$}}
\put(25.00,4.00){\makebox(0,0)[ct]{$4$}}
\put(11.00,25.00){\makebox(0,0)[rc]{$\mu$}}
\put(26.00,28.00){\makebox(0,0)[lb]{$\epsilon$}}
\put(19.00,24.00){\makebox(0,0)[lb]{$\gamma$}}
\put(27.00,17.00){\makebox(0,0)[ct]{$\nu$}}
\put(17.00,9.00){\makebox(0,0)[rt]{$\alpha$}}
\put(31.00,11.00){\makebox(0,0)[rt]{$\beta$}}
\put(15.00,35.00){\line(1,-3){10.00}}
\end{picture}
\end{center}
\caption{Spin~2}
\end{figure}
\begin{cor}[\rm spin 2]
Consider an arbitrary state of spin 2. Let $\ep$, $\mu$, $\nu$, $\al$, $\be$,
$\ga$
be the squares of the chordal distances between the four points on
the sphere of radius $\12$ that represent this state
(see figure). It's Wehrl entropy
is given by
\eq
S_W(\ep,\mu,\nu,\al,\be) = \frac{4}{5} + c \cdot \left( \sigma + \frac{1}{c}
\ln\frac{1}{c}\right), \label{S2}
\en
where
\eq
\frac{1}{c} = 1 - \frac{1}{4}\sum\lipic
+\frac{1}{12}\sum\papic \label{c2}
\en
and
\eq
\sigma = \frac{1}{12}\left(-\frac{1}{2}\sum\trpic
-\frac{5}{3}\sum\papic-\sum\wepic+3\sum\lipic
\right)
\en
with
\eq
\sum\trpic \equiv \al\mu\nu+\ep\be\nu+\ep\mu\ga+\al\be\ga,
\en
\eq
\sum\papic \equiv \al\ep+\be\mu+\ga\nu,
\qquad
\sum\lipic \equiv \al+\be+\ga+\mu+\nu+\ep,
\en
\eq
\sum\wepic \equiv \al\mu+\al\nu+\mu\nu+\be\ep
+\be\nu+\ep\nu+\ep\ga+\ep\mu+\mu\ga
+\al\be+\al\ga+\be\ga.
\en \label{spin2}
\end{cor}

\noindent \emph{Remark:} The fact that the four points lie on the surface
of a sphere imposes a complicated constraint on the parameters
$\ep$, $\mu$, $\nu$, $\al$, $\be$,
$\ga$. Although we have convincing numerical evidence
for Lieb's conjecture for spin~2,
so far a rigorous proof has been limited to
certain symmetric configurations
like equilateral triangles with centered fourth point ($\ep=\mu=\nu$ and
$\al=\be=\ga$), and squares ($\al=\be=\ep=\mu$ and
$\ga=\nu$). It is not hard to find values of the parameters
that give values of $S_W$ below the entropy for coherent states, but they
do \emph{not} correspond to any configuration of points on the sphere,
so in contrast to spin 1 and spin 3/2
the constraint is now important.
$S_W$ is concave in each of the parameters $\ep$, $\mu$, $\nu$, $\al$, $\be$,
$\ga$.

\noindent {\sc Proof}: The proof is analogous to the spin~1 and spin~3/2
cases but the geometry and algebra are considerably more complicated,
so we will just give a sketch. Pick four points on the sphere,
without loss of generality parametrized as $\om_1^{(3)} =(\tilde\theta,0)$,
$\om_2^{(3)} =(\theta,\phi)$, $\om_3^{(3)} = (0,0)$,
and $\om_4^{(3)} =(\bar\theta,\bar\phi)$. Corresponding spin $\12$ states
are $|\om_1^{(3)}\ra$, $|\om_2^{(3)}\ra$, $|\om_3^{(3)}\ra$,
as given in (\ref{om1}), (\ref{om2}), (\ref{om3}), and
\eq
|\om_4^{(3)}\ra = (1-\ga)^\12 e^{-i\bar\phi \over 2} |\U\ra
+ \ga^\12 e^{i\bar\phi \over 2} |\D\ra.
\en
Up to normalization, the state of interest is
\eq
|\tilde\psi^{(3)}\ra =
P_{j=2} |\om_1^{(3)} \ot \om_2^{(3)} \ot\om_3^{(3)} \ot\om_4^{(3)}\ra.
\en
In the computation of $|\tilde\psi^{(3)}_m|^2$ we encounter
again the angle $\phi$, compare (\ref{phi}),and two new
angles $\bar\phi$ and
$\bar\phi -\phi$.
Luckily both can again be expressed as angles between planes that
intersect the circle's center and we have
\eqa
2 \cos\bar\phi\sqrt{\mu\ga(1-\mu)(1-\ga)} & = & 
\mu + \ga - \al - 2\mu\ga, \\
2 \cos(\bar\phi-\phi)\sqrt{\ep\ga(1-\ep)(1-\ga)}
& = & \ga + \ep - \be - 2\ga\ep,
\ena
and find $1/c = \sum_m |\tilde\psi^{(3)}_m|^2$ as given in (\ref{c2}).
By permuting the parameters $\ep$, $\mu$, $\nu$, $\al$, $\be$,
$\ga$ appropriately we can derive expressions for the remaining
$|\tilde\psi^{(i)}_m|^2$'s and then compute $S_W$ (\ref{S2}) with the
help of $(\ref{formula})$.$\blob$

\section{Higher spin}

The construction outlined in the proof of corollary~\ref{spin2}
can in principle also be applied to states of higher spin, but
the expressions pretty quickly become quite unwieldy.
It is, however, possible to use theorem~\ref{theorem} to show that
the entropy is extremal for coherent states:

\begin{cor}[\rm spin $j$]
Consider the state of spin $j$ characterized by $2j -1$ coinciding
points on the sphere and a $2j$'th point, a small (chordal) distance
$\ep^\12$ away from them. The Wehrl entropy of this small deviation
from a coherent state, up to third order in $\ep$, is
\eq
S_W(\ep) = {2j\over 2j+1} + {c \over 8 j^2} \ep^2 \quad + {\cal O}[\ep^4] ,
\en
with
\eq
{1 \over c} = 1 - {2j - 1 \over 2j} \ep \quad \mbox{(exact)} .
\en
\end{cor}

A generalized version of Lieb's conjecture, analogous to (\ref{W2}), 
is \cite{A}
\begin{conj} \label{conject2}
Let $|\psi\ra$ be a normalized state of spin $j$, then
\eq
(2j s + 1) \int\dOm \, |\la\Om|\psi\ra|^{2s} \leq 1 , \quad s > 1 ,
\label{norms}
\en
with equality if and only if $|\psi\ra$ is a coherent state.
\end{conj}

\noindent \emph{Remark:} 
This conjecture is equivalent to the ``quotient of two H\"older inequalities"
(\ref{holder}).
The original conjecture~\ref{conject1} 
follows from it in the limit $s \rightarrow 1$.
For $s=1$ we simply get the norm of the spin $j$ state $|\psi\ra$,
\eq
(2j + 1) \int\dOm \, |\la\Om|\psi\ra_j|^2 = | P_j | \psi\ra|^2 ,
\label{norm}
\en
where $P_j$ is the projector onto spin $j$.
We have numerical evidence for low spin
that an analog of conjecture~\ref{conject2}
holds in fact for a much larger class of convex functions than
$x^s$ or $x \ln x$.

For $s \in \N$ there is a surprisingly simple group theoretic argument
based on (\ref{norm}):
\begin{thm}
Conjecture~\ref{conject2} holds for $s \in \N$. \label{natural}
\end{thm}

\noindent \emph{Remark:} For spin 1 and spin 3/2  (at $s=2$) this was
first shown by Wolfgang Spitzer by direct computation of the
integral.

\noindent {\sc Proof}: Let us consider
$s=2$, $|\psi\ra \in [j]$ with $||\psi\ra|^2 = 1$,
rewrite (\ref{norms}) as follows
and use (\ref{norm})
\eqa
\lefteqn{(2j\cdot 2 + 1) \int\dOm \, |\la\Om|\psi\ra|^{2\cdot 2}} \nn
&& = (2(2j)+ 1) \int\dOm \, |\la\Om\ot\Om|\psi\ot\psi\ra|^2 
=  |P_{2j} |\psi\ot\psi\ra |^2.
\ena
But $|\psi\ra\ot|\psi\ra \in [j]\ot[j] = [2j]\oplus[2j-1]\oplus\ldots
\oplus[0]$, so $|P_{2j} |\psi\ot\psi\ra |^2 < ||\psi\ot\psi\ra |^2 = 1$
unless $|\psi\ra$ is a coherent state, in which case
$|\psi\ra\ot|\psi\ra \in [2j]$ and we have equality. The proof for
all other $s \in \N$ is completely analogous.$\blob$\\[1ex]
It seems that there should also be a similar group theoretic
proof for all real, positive $s$ related to (infinite dimensional)
spin~$js$ representations of su(2) (more precisely: sl(2)). 
There has been some progress and it is now clear that there will
not be an argument as simple as the one given above \cite{J}.
Coherent states of the form discussed in \cite{C3} (for the hydrogen atom)
could be of importance here, since they easily generalize to non-integer
`spin'.

Theorem~\ref{natural} provides a quick, crude, lower limit on the
entropy:

\begin{cor}
For states of spin $j$
\eq
S_W(|\psi\ra\la\psi|) \geq \ln{4j+1\over 2j+1} > 0.
\en
\end{cor}

\noindent {\sc Proof}: This follows from Jensen's inequality and
concavity of $\ln x$:
\eqa
S_W(|\psi\ra\la\psi|) & = &
-{\textstyle (2j+1)}\int\dOm\, |\la\Om|\psi\ra|^2 \ln |\la\Om|\psi\ra|^2 \nn
& \geq & -\ln\left({\textstyle (2j+1)} 
\int\dOm\, |\la\Om|\psi\ra|^{2\cdot 2}\right) \nn
& \geq & -\ln{2j+1 \over 4j + 1} .
\ena
In the last step we have used theorem~\ref{natural}.$\blob$\\[1ex]
We hope to have provided enough evidence to convince the
reader that it is reasonable to expect that
Lieb's conjecture is indeed true for all spin. All cases
listed in Lieb's original article, $1/2$, $1$, $3/2$, are now
settled --
it would be nice if someone
could take care of the remaining ``dot, dot, dot" \ldots 

\section*{Acknowledgments}

I would like to thank Elliott Lieb 
for many discussions, constant support, encouragement, and for reading the
manuscript.
Much of the early work was done in collaboration with Wolfgang Spitzer.
Theorem~\ref{natural} for spin 1 and spin 3/2 at $s=2$ is due to him and
his input was crucial in eliminating many other plausible approaches.
I would like to thank him for many discussions and excellent team work.
I would like to thank Branislav Jur\v co for joint work on the
group theoretic aspects of the problem and stimulating discussions
about coherent states.
It is a pleasure to thank Rafael Benguria, Almut Burchard, Dirk Hundertmark,
Larry Thomas, and Pavel Winternitz for many valuable discussions.
Financial support by the Max Kade Foundation is gratefully acknowledged.

\end{document}